\documentclass[12pt]{article}
\voffset - 1 in
\usepackage{amsfonts}
\usepackage{amsmath}
\usepackage{amssymb,amsbsy}
\usepackage{bm}
\usepackage{makeidx}
\usepackage{latexsym}
\usepackage{graphicx}
\usepackage{epsfig}
\usepackage{subfigure}
\usepackage{dsfont}
\usepackage{mathcomp}
\usepackage{color}
\usepackage{mathrsfs}
\usepackage{epsf}
\usepackage{authblk}
\newcommand{\beq}{\begin{eqnarray}}% can be used as {eqnarray}
\newcommand{\eeq}{\end{eqnarray}}
\newcommand{\be}{\begin{equation}}% can be used as {equation} 
\newcommand{\ee}{\end{equation}}

\begin{document}

\title{\bf Rescuing Quadratic Inflation \\ ~~}
\date{}
\author[a,b]{{\bf John Ellis}\thanks{john.ellis@cern.ch}}
\author[a]{{\bf Malcolm Fairbairn}\thanks{malcolm.fairbairn@kcl.ac.uk}}
\author[a]{{\bf Maria Sueiro}\thanks{maria.sueiro@kcl.ac.uk}}
\affil[a]{\small Theoretical Particle Physics and Cosmology Group, Department of Physics, \newline King's College London, Strand, London WC2R 2LS, UK}
\affil[b]{\small Theory Division, Physics Department, CERN, CH-1211, Geneva 23, Switzerland}

\renewcommand\Authands{ and }

\maketitle
\begin{abstract}
Inflationary models based on a single scalar field $\phi$ with a quadratic potential $V = \frac{1}{2} m^2 \phi^2$
are disfavoured by the recent Planck constraints on the scalar index, $n_s$, and the tensor-to-scalar ratio
for cosmological density perturbations, $r_T$. In this paper we
study how such a quadratic inflationary model can be rescued by postulating additional fields with
quadratic potentials, such as might occur in sneutrino models, which might serve as either
curvatons or supplementary inflatons. Introducing a second scalar field reduces but does not
remove the pressure on quadratic inflation, but we find a sample of three-field models that
are highly compatible with the Planck data on $n_s$ and $r_T$. We exhibit a specific three-sneutrino example that is
also compatible with the data on neutrino mass difference and mixing angles.\\

\begin{center}
KCL-PH-TH/2013-40, LCTS/2013-27, CERN-PH-TH/2013-293
\end{center}

\end{abstract}

\section{Introduction}

Inflation is a very promising paradigm for the behaviour of the scale factor in the early Universe,
which offers a solution to the cosmological horizon problem, explaining the observed large-scale isotropy
and giving rise to small fluctuations in energy density as observed in the cosmic microwave
background (CMB) radiation that seeded galactic structure formation, as well as
explaining the absence of topological defects.  In the simplest models of inflation, 
the energy density responsible for the accelerated early expansion of the scale factor
arises from the expectation value of a scalar field rolling down a potential.
As discussed in~\cite{EI}, a huge number of candidate scalar fields appearing in extensions of the standard model
of particle physics and in models with modified gravitational sectors
have the freedom to match observations: for a sampling of recent models,
see~\cite{CEM,EKN,NTY,KL,Buch,Farakos:2013cqa,Roest:2013aoa,K,Ellis:2013zsa,FS,LLN}.

In the simplest class of single-field models, the energy density in the potential
causes an accelerated expansion of the Universe, with the strength of tensor perturbations
being directly related to the magnitude of the energy density. Since the inflaton $\phi$ rolls slowly down the potential,
the spectrum of scalar perturbations is tilted. Once the value of $\phi$ becomes sub-Planckian, 
the slow roll ends and the field starts to oscillate and then decays into radiation, reheating the Universe.
This simplest model makes surprisingly successful predictions for the main inflationary observables.
For example, it provides an almost scale-invariant spectrum of density perturbations 
and predicts that the non-Gaussianity parameter $f_{NL}$ is relatively small. 

The simplest scenario is a single scalar field $\phi$ with a monomial potential such as
$\frac{1}{2}m^2\phi^2$ or $\lambda\phi^4$. However, $\lambda\phi^4$ models were already under
pressure from the results of WMAP~\cite{WMAP} and can now be regarded as excluded by
the observations of the Planck satellite~\cite{planck,planckinflation},
and quadratic $\frac{1}{2}m^2\phi^2$ models are now also under severe pressure, 
since they are unable to fit simultaneously the observed 
spectral index of scalar perturbations $n_s$ and the low tensor-to-scalar ratio $r_T$.

For a range of typical values of the 
number of efolds during inflation, $N_*\in[40,60]$, single-field inflationary models with a
$\frac{1}{2}m^2\phi^2$ potential predict a value for $r_T$ in the range $[0.1,0.2]$, 
whereas the Planck data require $r_T < 0.1$. For a lower value of $N_*$, i.e.,
a smaller amount of expansion between the largest scales leaving the horizon and the end of inflation, 
it is possible for a quadratic model to give a value for $r_T$ that obeys the Planck constraint, 
but the quadratic model then makes a prediction for the spectral index of the perturbations
$n_s$ that does not respect the Planck constraint $n_s=0.9624 \pm 0.0075$, since there is
a one-to-one mapping between the number of efolds and the scalar spectral index.

In this work we explore ways to rescue models with quadratic potentials.
Our objective is to find a model of quadratic inflation that (i) gives the correct magnitude for the
scalar density perturbation $\delta$, (ii) yields a Planck-compatible value of the scalar
spectral index $n_s$, (iii) predicts a value for the tensor-to-scalar ratio $r_T$ that obeys the 
Planck constraint, and (iv) does not predict significant non-Gaussianity. 

Attaining all these
objectives simultaneously requires modifying the mapping between the number of efolds 
and the scalar spectral index. This may be done by introducing extra fields that act either
as secondary inflatons or as curvatons. We restrict our attention to minimal models with
multiple $\frac{1}{2}m^2\phi^2$ potentials. Apart from simplicity, one of our motivations
for this restriction is the idea that the inflaton might be a singlet sneutrino in a
supersymmetric seesaw model of neutrino masses~\cite{ERY}, which is a motivated and minimal
extension of the Standard Model. In such a scenario, there must be at least two such
sneutrino fields so that two light neutrinos can acquire masses, and probably three
so that all three light neutrinos can be massive.

In the next Section we review the simplest case of a single $\frac{1}{2}m^2\phi^2$ 
inflationary field, and the above-mentioned reasons why it has difficulties fitting the latest data.
In Section~3 we go on to examine models with an additional scalar field acting as a curvaton.
Whilst the disagreement with the data can be reduced in such scenarios, they are
still in some tension with the data, although not enough to be considered as excluded. Then, in Section~4
we look at models with two inflatons acting in consort accompanied by an curvaton, which we find
to be the minimal combination of quadratic fields that obeys completely all the Planck constraints. 
In Section~5 we show how such a supersymmetric seesaw model can be realized with the
three quadratic scalar fields being interpreted as supersymmetric partners of heavy singlet neutrinos,
showing how such a model can fit simultaneously the available data on neutrino
oscillations. Finally, Section~6 contains a summary and discusses future prospects
for quadratic inflationary models.

\section{Single-Field Quadratic Inflation\label{infsec}}

We first review the standard lore for single-field inflation with a
quadratic potential $V=\frac{1}{2}m_\phi^2\phi^2$.  The power spectrum of the scalar perturbations is given 
by~\cite{liddlelyth1,Byrnes:2006fr}
\be\label{eq:spectrum1}
\mathcal{P}_\zeta = \left(\frac{H_*}{\dot{\phi_*}}\right)^2\left(\frac{H_*}{2\pi}\right)^2 = \frac{\phi_*^2}{4 M_{Pl}^4}\left(\frac{H_*}{2\pi}\right)^2 \, ,
\ee
where starred quantities correspond to the values of the various parameters at the epoch 
at which the last scales to re-enter the horizon before the last scattering surface were 
initially leaving the horizon during inflation.  The Hubble parameter is given by the Friedmann 
equation $H^2=V/(3M_{Pl}^2)$, where $M_{Pl}$ is the reduced Planck mass, 
$M_{Pl}=(8\pi G)^{-1/2}=2.4\times 10^{18}$~GeV.  The density perturbation is 
$\delta_H^2\equiv \frac{4}{25}\mathcal{P}_\zeta$ ~\cite{liddlelyth1}, so using the above equation we have
\be
\delta_H=\sqrt{\frac{1}{600 \pi^2}}\frac{m_\phi\phi_*^2}{M_{Pl}^3} .
\ee
The two slow-roll parameters in this case are:
\be
\epsilon \equiv \frac{M_{Pl}^2}{2}\left(\frac{V'}{V}\right)^2 = \frac{2M_{Pl}^2}{\phi^2}\qquad ;\qquad
\eta \equiv M_{Pl}^2\frac{V''}{V} = \frac{2 M_{Pl}^2}{\phi^2} \, ,
\ee
where dashed quantities denote derivatives with respect to $\phi$. 
 
The scalar spectral index is given in terms of the slow-roll parameters
by $n_s - 1 = -6 \epsilon_* + 2 \eta_*$~\cite{Wands:2002bn}, so in this case it is given by
\be
n_s-1 = -\frac{8M_{Pl}^2}{\phi_*^2} .
\ee
The tensor-to-scalar ratio $r_T$ is defined as the ratio of the power spectrum of the tensor perturbations~\cite{Liddle:1993fq}
\be
\mathcal{P}_T = \frac{8}{M_{Pl}^2}\left(\frac{H_*}{2\pi}\right)^2
\ee
to the power spectrum of the scalar perturbations, given in (\ref{eq:spectrum1}).
In the simple quadratic inflation model the tensor-to-scalar ratio is given by
\be
r_T \equiv \frac{\mathcal{P}_T}{\mathcal{P}_\zeta} = \frac{32 M_{Pl}^2}{\phi_*^2} \, ,
\ee
yielding a very restrictive relationship between the tensor-to-scalar ratio and the scalar spectral index:
\be
r_T=4(1-n_s) \, .
\label{rigid}
\ee
Ultimately, both quantities are fixed by the expectation value of the inflaton
corresponding to the scales entering the horizon at the last scattering surface, 
which is around 40-60 efolds before the end of inflation.  
In a particular model the exact number of efolds is not a free parameter,
but depends upon the rate at which the coherent oscillations of the inflaton decay into radiation after inflation.

To find the time $t_*$ when the largest scale crosses the horizon during inflation,
we must solve the equation
\be\label{eq:efolds}
N_*=60 -\ln \frac{10^{16}~{\rm GeV}}{V_*^{1/4}}+\ln\frac{V_*^{1/4}}{V_{end}^{1/4}}-\frac{1}{3}\ln\frac{V_{end}^{1/4}}{\rho_{\textrm{last decay}}^{1/4}} \, ,
\ee
where $V_{end}$ is the potential at the end of inflation and $\rho_{\textrm{last decay}}$ 
is the energy density evaluated at the time of the decay of the longest-lived field in the theory ~\cite{liddlelyth1}. 
Solving (\ref{eq:efolds}) is quite simple for quadratic inflation, and means that the number of efolds $N_*$,
and therefore the time of horizon crossing $t_*$, is not an arbitrary parameter but is constrained for a 
given set of masses and decay rates. A change in the duration of matter or radiation domination of the 
energy density up until the last decay when the Universe is thermalized for the last time will affect the value of $N_*$.
In the case of single-field $\phi^2$ (quadratic) inflation, the two model parameters are
therefore the inflaton mass $m_\phi$ and its decay rate $\Gamma_\phi$, as shown in 
the first row of Table~\ref{tab:parameters}.

\begin{table}
\begin{center}
\begin{tabular}{ |c|c|}
\hline
$m_\phi$ & $\Gamma_\phi$ \\
\hline
\end{tabular} \\
%\end{center}
%\caption*
\vspace{0.3cm}
{ \it Free parameters for single-field quadratic inflation.} \\
%\begin{center}
\vspace{0.3cm}
\begin{tabular}{|c|c|c|c|c|}
\hline
$m_\phi$ & $\Gamma_\phi$ & $m_\sigma$ & $\Gamma_\sigma$ & $\sigma_*$ \\
\hline
\end{tabular} \\
\vspace{0.3cm}
%\end{center}
%\caption
{ \it Free parameters for the model with a quadratic inflaton and a quadratic curvaton.}\\
\vspace{0.3cm}
%\begin{center}
\begin{tabular}{|c|c|c|c|c|c|c|}
\hline
$m_\phi$& $m_\chi$& $\Gamma_\chi$& $\theta_{\phi\chi*}$& $m_\sigma$& $\Gamma_\sigma$& $\sigma_*$ \\
\hline
\end{tabular} \\
\vspace{0.3cm}
%\caption
{ \it Free parameters for the model with two quadratic inflatons and a quadratic curvaton.}\\
\end{center}
\caption{Free parameters in the quadratic inflation models discussed in the text.}
\label{tab:parameters}
\end{table}

\begin{figure}[!htbp]
\centering
\includegraphics[scale=0.6]{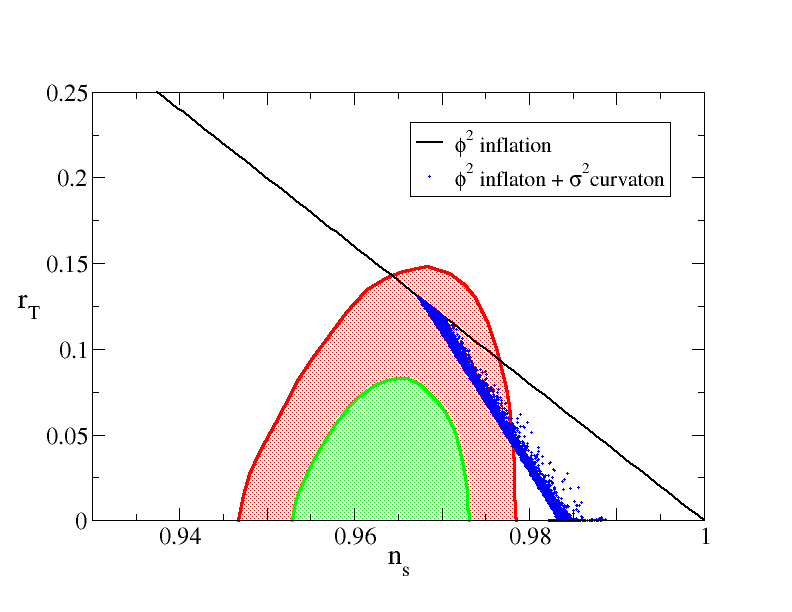}
\caption{\it The plane of the spectral index $n_s$ and the tensor-to-scalar 
ratio $r_T$. The shaded regions correspond to the 1- and 2-$\sigma$ constraints obtained by
combining  Planck and baryon acoustic oscillation (BAO) data~\cite{planckinflation}.  
The single black line corresponds to minimal $\phi^2$ (quadratic) inflation,
and the blue points are combinations of a $\phi^2$ inflaton and a $\sigma^2$ (quadratic)
curvaton, which reduces to $\phi^2$ inflation as a limiting case when the curvaton corrections are negligible.}
\label{curvplanck}
\end{figure}

The rigid relationship (\ref{rigid}) between $n_s$ and $r_T$ is plotted in Fig.~\ref{curvplanck},
where it can be seen that single-field $\phi^2$ (quadratic) inflation does not fit the data very well, 
no matter what the number of efolds.

\section{Model with one Inflaton and one Curvaton}

We now consider a second model that contains, in addition to the field $\phi$ with a quadratic potential that
acts as the inflaton, a second scalar field with a quadratic potential that acts as a the curvaton $\sigma$~\cite{Enqvist:2001zp, Lyth:2002my, Moroi:2005kz, Ichikawa:2008iq}.
This model is therefore described by 
\be
V=\frac{1}{2}m_\phi^2\phi^2+\frac{1}{2}m_\sigma^2\sigma^2 \, .
\ee
The curvaton field $\sigma$ is frozen at an expectation value $\sigma_*$ during the slow-roll of $\phi$
since $H\gg m_\sigma$. After inflation ends and the Hubble parameter becomes comparable to the 
curvaton mass, $H\sim m_\sigma$, the curvaton begins to oscillate around the minimum of its potential. 
When the Hubble parameter falls to the curvaton decay rate, $\Gamma_\sigma$, 
the curvaton decays to radiation. The parameters of this model are the masses and decay rates of the 
inflaton and the curvaton, $m_\phi$, $\Gamma_\phi$ and $m_\sigma$, $\Gamma_\sigma$ respectively, 
and the expectation value of the curvaton during inflation $\sigma_*$, as listed in the second
row of Table~\ref{tab:parameters}.

\begin{figure}[!htbp]
\centering
\begin{tabular}{cc}
\includegraphics[width=80mm, height=80mm]{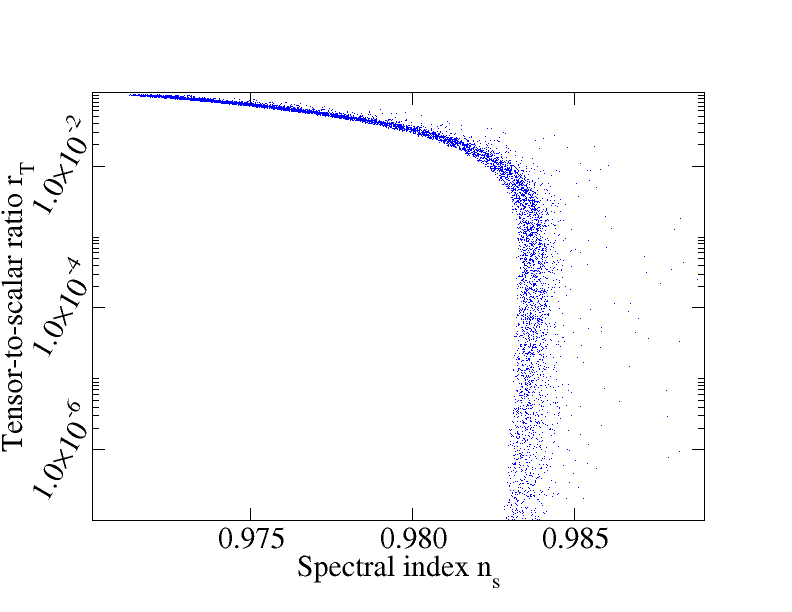}&
\includegraphics[width=80mm, height=80mm]{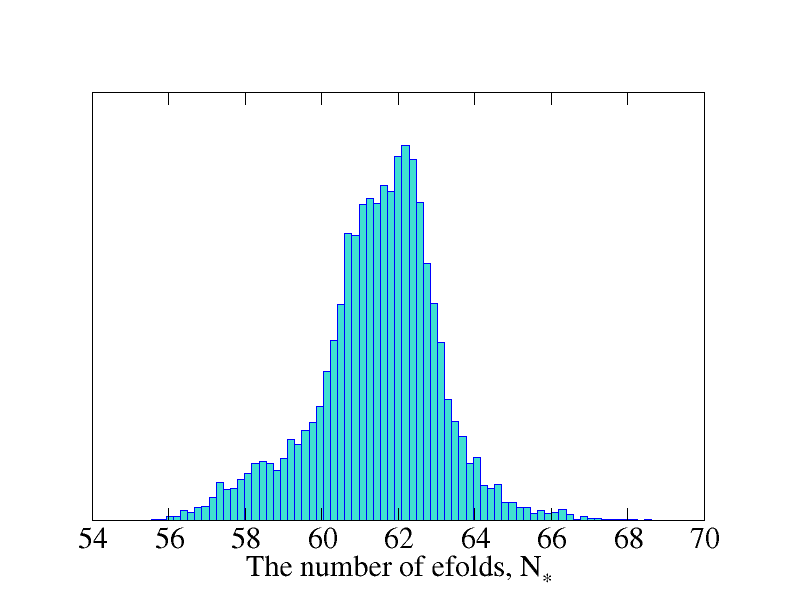}\\
\end{tabular}
\caption{\it The left panel shows the spectral index $n_s$ and the tensor-to-scalar ratio $r_T$ in
the model with one inflaton and one curvaton (same data as figure \ref{curvplanck} but on log scale). Here, we show only the points that have a value for $r_T$ 
that is lower than the Planck constraint of $r_T<0.1$. This occurs only for values of $n_s$ that are 
significantly larger than that measured. In the right panel, we show the values of the number 
of efolds $N_*$ in this model, which is in the range $[55,70]$.}
\label{fig:rtns-efolds1}
\end{figure}

Now, although only $\phi$ is responsible for providing the expansion of the Universe during inflation, 
both fields can contribute to $\delta_H$, $n_s$, and $r_T$. From ~\cite{Lyth:2001nq, Dimopoulos:2003ss}
we see that the total density perturbation is given by
\be\label{eq:denspert1}
\delta_H = {\delta_H}_\phi + \frac{f_\sigma}{3}{\delta_H}_\sigma \, ,
\ee
with the individual components evaluated at horizon crossing~\footnote{All quantities 
evaluated at this time have the index *.}, where we have
\be\label{eq:denspertind}
{\delta_H}_\phi = \sqrt{\frac{1}{600 \pi^2}}\frac{m_\phi \phi_*^2}{M_{Pl}^3}\qquad ; \qquad
{\delta_H}_\sigma =\frac{2}{5}\left(\frac{H_*}{\pi \sigma_*}\right) \, ,
\ee
and $f_\sigma$ is evaluated at the epoch of the last decay, i.e., the epoch 
at which the curvaton decays, rather than the inflaton, so that
\be
f_\sigma=\left(\frac{3 \Omega_\sigma}{4-\Omega_\sigma}\right)_{\textrm{last decay}} \, ,
\ee
where $\Omega_\sigma=\rho_\sigma/\rho_{tot}$ at the epoch of curvaton decay ~\cite{Enqvist:2008gk,Fonseca:2012cj,Sueiro:2012fq}.  
The power spectrum of the created density perturbation is then
\be\label{eq:spectrum}
\mathcal{P}_\zeta \backsim {\delta_H}_\phi^2+\frac{f_\sigma^2}{9}{\delta_H}_\sigma^2 \, ,
\ee
and to evaluate the spectral index we simply use the definition
\be\label{eq:spectral}
n_s-1 \equiv \frac{d\ln \mathcal{P}}{d\ln k} = \frac{\sqrt{3}M_{Pl}}{\sqrt{V_*}\mathcal{P}}\frac{d\mathcal{P}}{dt} \, ,
\ee
where we transformed the derivative using $d\ln k = H dt$, 
and note that the expansion $H$ is dominated by the effects of $V$ at this epoch, 
which is during slow-roll inflation.  The tensor-to-scalar ratio is
\be\label{eq:ttos}
r_T\equiv\frac{\mathcal{P}_T}{\mathcal{P}_\zeta} \, ,
\ee
which may be evaluated using (\ref{eq:spectrum}) and
\begin{equation}
\mathcal{P}_T \; = \; \frac{2V_*}{3\pi^2M_{Pl}^4} \, .
\end{equation}
Once the time $t_*$ has been established in the same way as in Section~\ref{infsec}, 
we can calculate the density perturbation using (\ref{eq:denspert1}), the spectral index using (\ref{eq:spectral}), 
and the tensor-to-scalar ratio using (\ref{eq:ttos}). Employing a Markov Chain Monte Carlo algorithm,
we fit the measured values of $\delta_H$ and $n_s$ and explore the predictions for $r_T$,
as shown in Figs.~\ref{curvplanck} and \ref{fig:rtns-efolds1}. 

We find in our sampling of its five-parameter space that
this model is able to predict a low value for the tensor-to-scalar ratio $r_T$, well under the constraint coming from
Planck. This happens when the curvaton field is significant at the time of the last decay,
and dominates the energy density. However, obtaining a low value for $r_T$ is possible only
at the expense of an unsuccessful prediction for the scalar spectral index. As is shown in the left panel of 
Fig. (\ref{fig:rtns-efolds1}), when $r_T$ falls to a value below $0.1$, 
$n_s$ grows to a value that is approximately $3\sigma$ away from the current measurement
as $r_T \to 0$.

This conclusion is qualitatively the same as the results displayed in fig. 6 of ~\cite{Moroi:2005np} and fig. 1 of ~\cite{Enqvist:2013paa}. In both cases it was shown that in a model with one inflaton and one curvaton, a low predicted value for $r_T$ is possible but it leads to a higher value of $n_s$. However, that analysis was done for a set range of the number of efolds $N$ as well as a set value of the curvaton's contribution to the perturbations.  Our results, shown in Figs.~\ref{curvplanck} and \ref{fig:rtns-efolds1}, differ from the results in~\cite{Fonseca:2012cj}, where a Planck-consistent value of $r_T$
was found in a fit to the values of $\delta_H$ and $n_s$ in a model with one inflaton and one curvaton. 
The reason for this difference is that in~\cite{Fonseca:2012cj} the inflationary scale at horizon crossing
was set by hand by assuming $\epsilon_*=0.02$, which corresponds to a very low number of efolds, $N_*\backsim 25$. 
In contrast, in our work we solve for the field value at $t_*$, so that our model is self-consistent. 
We find that $N_*$ has to be in the range $[55,70]$ as shown in the right plot of Fig. (\ref{fig:rtns-efolds1}).

\begin{figure}[!htbp]
\centering
\begin{tabular}{cc}
\includegraphics[width=80mm, height=80mm]{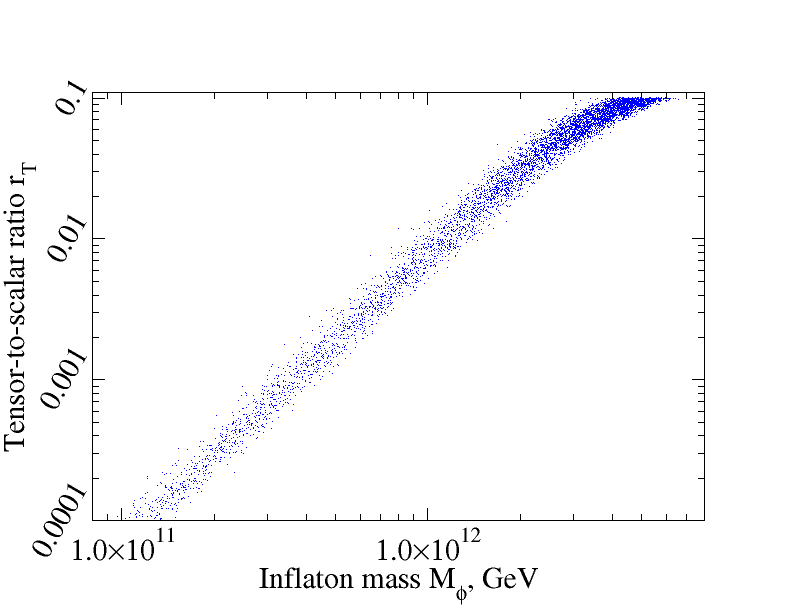}&
\includegraphics[width=80mm, height=80mm]{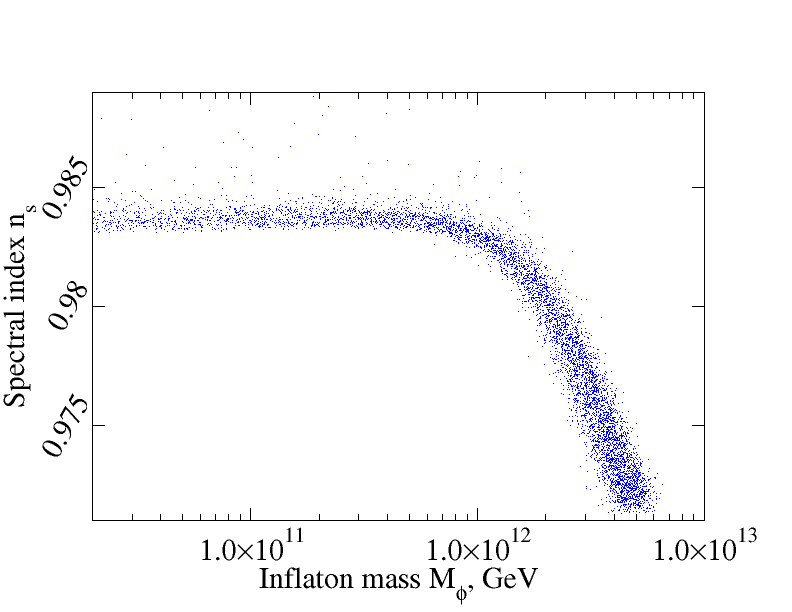}\\
\end{tabular}
\caption{\it The left panel shows the dependence of the tensor-to-scalar ratio $r_T$ on the inflaton mass $m_\phi$. 
We see that obtaining a lower value for $r_T$, so as to be consistent with the constraint coming from
Planck, would require a lower inflaton mass. However, as is shown in the right panel, 
lower values of $m_\phi$ result in values of $n_s$ that are significantly higher than the measured one.}
\label{fig:rtnsmphi1}
\end{figure}

We find that the values of $r_T$ and $n_s$ depend strongly on the value of the inflaton mass, $m_\phi$.
 As is shown in the left panel of Fig.~\ref{fig:rtnsmphi1}, 
 a lower value of $r_T$ can be obtained with a lower value of $m_\phi$, 
 but this corresponds to a higher value of $n_s$. The same is observed in fig. 6 and 7 of ~\cite{Meyers:2013gua}; for an analytic solution, the authors study $n_s$ and $r_T$ in terms of the curvaton's expectation value and the ratio of the decay rates of the inflaton and the curvaton. They find that the combinations of parameters that result in a Planck-consistent value of $r_T$ are associated with a value of $n_s$ that is significantly higher than the measured one, which is what we find as well.
 This behaviour is seen also in Fig.~\ref{curvplanck},
 where one can see that, although curvaton models fit the Planck data somewhat better than single-field
 quadratic inflation, they are still disfavoured, as they do not come within the 68\% CL region
 favoured by the Planck + BAO data.

We therefore conclude that, although a model with one inflaton and one curvaton, 
with both fields contributing to the generation of the perturbations, 
is able to predict a value for $r_T$ that is within the Planck constraint,
this model cannot combine this good prediction with a
successful prediction for the spectral index $n_s$.

This conclusion could be relaxed by assuming additional inflation at some later 
time in the Universe as in, e.g., thermal inflation~\cite{liddlelyth1,Liddle:1993fq}. However,
it seems quite difficult to obtain enough thermal inflation to reconcile the
Planck data on $r_T$ and $n_s$ with the model combining a quadratic inflaton and a quadratic curvaton,
since typically one requires 25-30 efolds of later inflation.

\section{Model with two Inflatons and one Curvaton}

In view of this setback, in this Section we further augment the model by including 
another scalar field, $\chi$. This is assumed to be a second inflaton field, 
again with a quadratic term in the potential, so that it becomes
\be\label{eq:pot2}
V=\frac{1}{2}m_\phi^2\phi^2+\frac{1}{2}m_\chi^2\chi^2+\frac{1}{2}m_\sigma^2\sigma^2 \, .
\ee
In this case, both $\phi$ and $\chi$ roll slowly down their potentials, giving rise to inflation while,
as before, the curvaton field $\sigma$ remains frozen at its expectation value during inflation and 
then oscillates and decays into radiation.

Curvaton domination of the energy density in this model can lead to a low value of $r_T$
consistent with the Planck data, while the addition of the second inflaton helps to
lower the prediction for $n_s$, making the model consistent with the Planck value. 
This is because, in the case of curvaton dominance, the power spectrum of the
scalar density perturbations is simply $P_\zeta\backsim(H_*/(\pi \sigma_*))^2$. 
Then, from the expression (\ref{eq:spectral}) for the spectral index we obtain the expression:
\be\label{eq:spectralcurv}
n_\sigma = \frac{\sqrt{3}M_{Pl}}{V_*^{3/2}}\frac{dV_*}{dt}+1 \, .
\ee
We see from (\ref{eq:spectralcurv}) that, in order to alter the high value of $n_s$ that was 
obtained in the two-field $(\phi, \sigma)$ model we must alter the time derivative of the 
potential evaluated at horizon crossing. Since the curvaton expectation value does not 
change during inflation, it is only with the addition of the second inflaton $\chi$ that we can achieve this. 

Substituting (\ref{eq:pot2}) in (\ref{eq:spectralcurv}), we obtain the following expression 
for the scalar spectral index in terms of our model parameters:
\be
n_\sigma - 1 = \frac{-M_{Pl}^2m_\phi^4\phi_*^2}{V_*^2}\left(1+\left(\frac{m_\chi}{m_\phi}\right)^4\left(\frac{\chi_*}{\phi_*}\right)^2\right) \, .
\ee
In our numerical analysis, we find that the heavier of the two inflaton fields (let us call it $\phi$) 
stops rolling once its expectation value drops down to the Planck scale. 
It then begins to oscillate, but the Universe is still expanding exponentially due to the other inflaton field. 
This means that $\phi$ is redshifted, with its expectation value dropping rapidly, and by the end of inflation, 
i.e., when the expectation value of $\chi$ becomes equal to $M_{Pl}$, the contribution of $\phi$ to the 
energy density is negligible. It is therefore not necessary to include the decay rate of the 
heavy inflaton among the relevant parameters of our model listed in the third row of
Table \ref{tab:parameters}.  The parameter $\theta_{\phi\chi*} \equiv \tan^{-1}( \chi_* / \phi_* )$ corresponds to the 
angle in field space at $t_*$, i.e. the angle between the $\phi$ and $\chi$ direction when the 
largest scales in the CMB are leaving the horizon. Typically the lighter field $\chi$ will be frozen at this moment, 
although we do not assume this, since we analyze this stage numerically.

\begin{figure}[!htbp]
\centering
\begin{tabular}{cc}
\includegraphics[width=80mm, height=80mm]{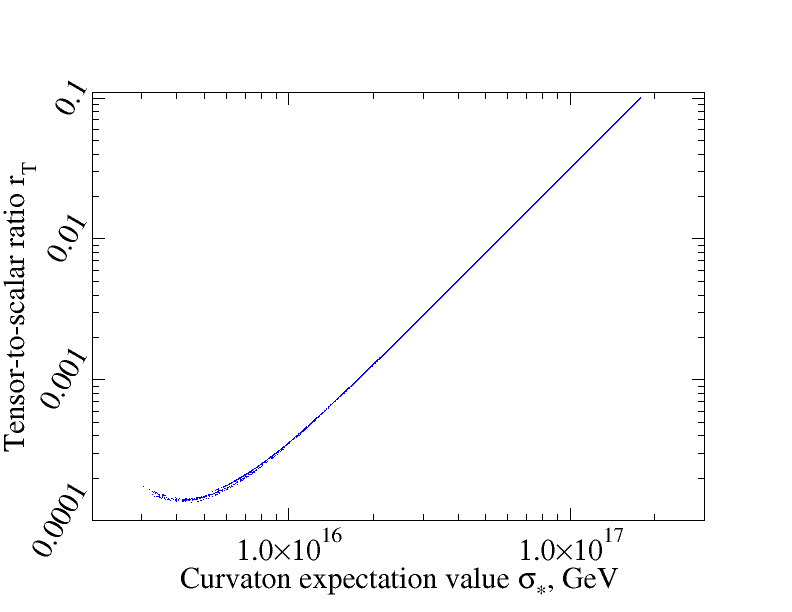}&
\includegraphics[width=80mm, height=80mm]{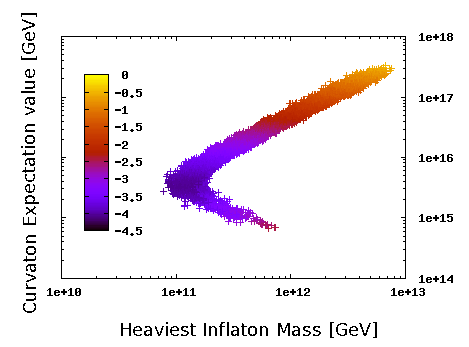}\\
\end{tabular}
\caption{\it The left panel shows the strong dependence of $r_T$ on the curvaton expectation value 
$\sigma_*$ in the model with two inflatons and one curvaton with quadratic potentials. 
In the right panel we plot the mass of the heaviest inflaton vs. the curvaton expectation value. 
The colours denote ranges of $\log_{10}(r_T)$.  In both plots the turn in the curve separates 
curvaton behaviour from pure inflationary behaviour in which the curvaton energy density always remains small.}
\label{fig:rtnssigma3}
\end{figure}

As in the model with one inflaton and one curvaton, we solve equation (\ref{eq:efolds})
numerically to find the time during inflation when the largest scales left the horizon and 
determine the expectation values $\phi_*$ and $\chi_*$ at that time. 
Since we are looking at the cases when the curvaton energy density dominates the 
Universe at the time of its decay, we can compute the density perturbation from the 
curvaton component of (\ref{eq:denspert1}) and the spectral index from (\ref{eq:spectralcurv}). 
The prediction for the tensor-to-scalar ratio $r_T$ is again calculated from (\ref{eq:ttos}), 
but in this case the spectrum of scalar perturbations is given by the second term on the right-hand side of 
(\ref{eq:spectrum}):
\be\label{eq:spectrum3}
P_\zeta=\frac{f_\sigma^2}{9}\left(\frac{H_*}{\pi \sigma_*}\right)^2
\ee
where in the case of curvaton domination $f_\sigma$ will be very close to $1$. 
This results in the expression
\be\label{eq:rtcurv}
r_T=\frac{9}{f_\sigma^2}\frac{2\sigma_*^2}{M_{Pl}^2} 
\ee
for the tensor-to-scalar ratio.

As in the previous Section, we employ a Markov Chain Monte Carlo algorithm to study
the predictions of this model for $r_T$, while fitting the measured values of $\delta_H$ and $n_s$. 
We find that this model with two inflatons and one curvaton is indeed able to accommodate values of $r_T$ 
that are in agreement with the Planck results. The left panel of 
Fig.~\ref{fig:rtnssigma3} shows the spectral index and the tensor-to-scalar ratio of this 3-field model,
%We see that, in contrast to the 2-field model, it is possible to obtain a value of the spectral index that correctly ifts the measurement while keeping the value of the tensor-to-scalar ratio lower than the prediction from $\phi^2$ single field inflation.  
and the model is compared to the Planck data in Fig.~\ref{neut}.

\begin{figure}[!htbp]
\centering
\begin{tabular}{cc}
\includegraphics[width=80mm, height=80mm]{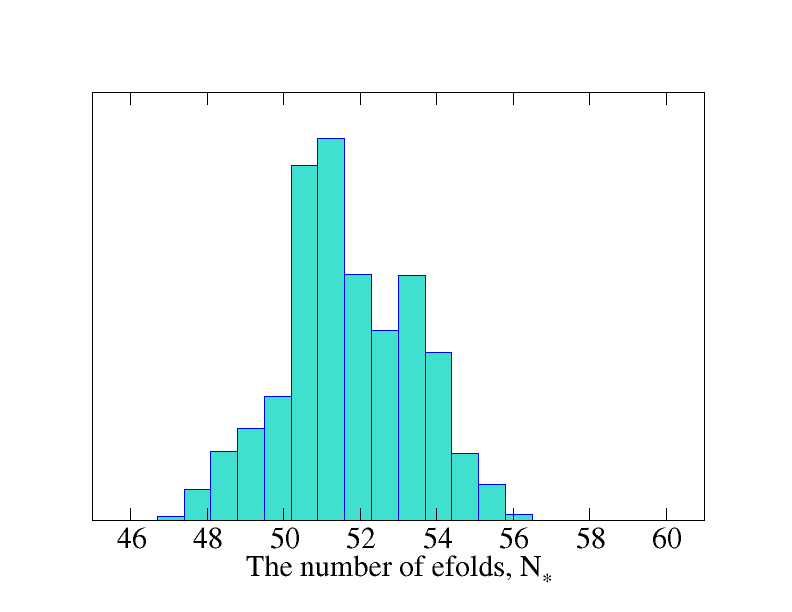}&
\includegraphics[width=80mm, height=80mm]{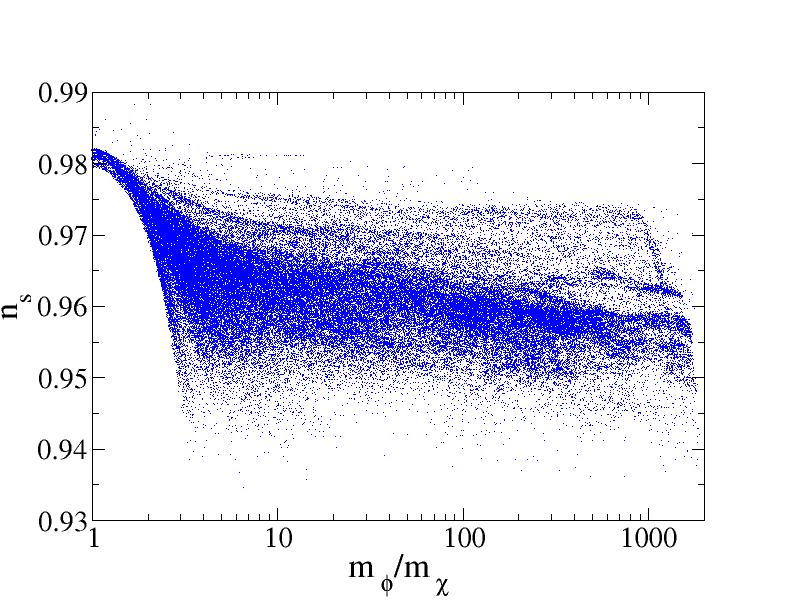}\\
\end{tabular}
\caption{\it The left panel shows the number of efolds in the model with 2 inflatons and a curvaton, 
which is slightly lower than for the curvaton scenario with a single inflaton.  
The right panel shows the dependence of the spectral index $n_s$ on the ratio of $m_\phi$ and 
$m_\chi$, showing that degenerate situations cannot lead to a low enough spectral index.\label{efspec}}
\label{fig:rtnssigma3} 
\end{figure}

\begin{figure}[!htbp]
\centering
\includegraphics[scale=0.6]{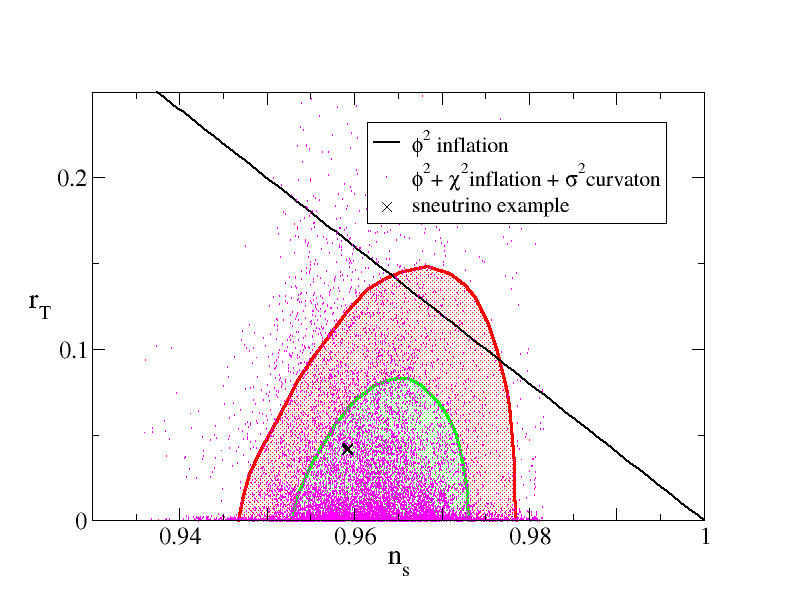}
\caption{\it The model with two inflatons and a curvaton (\protect\ref{eq:pot2}) yields good inflationary parameters.
Many of the mauve points lie within the region favoured by Planck at the 68\% CL. 
We also show as a black cross a good point in parameter space where the three fields are 
assumed to be sneutrinos and we also fit the neutrino mass differences and mixings.}
\label{neut}
\end{figure}

Fig.~\ref{fig:rtnssigma3} confirms what we expect from (\ref{eq:rtcurv}),
namely that a lower value of $r_T$ can be obtained from a lower curvaton expectation value $\sigma_*$. 
However, we also find that there is a minimum in the prediction for the value of $r_T$. 
As the curvaton expectation value becomes smaller, it becomes more difficult for the 
curvaton to dominate completely the energy density of the Universe at the time of its decay. 
Eventually, for $\sigma_*\backsim 5 \times 10^{15}$~GeV the parameter measuring the 
significance of the curvaton energy density, $f_\sigma$, becomes smaller than $1$.
We see in Fig.~\ref{fig:rtnssigma3} that this corresponds to a minimum value of $r_T$
around $3 \times 10^{-5}$. For even smaller curvaton expectation values, the curvaton energy 
density becomes less significant and  $f_\sigma$ rapidly becomes smaller. 
This means that the model becomes more and more like the single-field inflation model.
In this case we see that the value of $r_T$ begins to grow, as expected.  In Fig.~\ref{efspec} 
we see that the number of efolds $N_*$ is slightly lower than in the case with one inflaton and one curvaton,
and that in order to obtain a good spectral index we require $m_\phi$ to be greater than two or three times $m_\chi$.

It is essential that all the fields in our model decay before nucleosynthesis starts. 
This requirement places a lower constraint on the temperature $T_{reh}$ of the Universe when the last 
particle in our model decays into radiation, namely $T_{reh}>2$~MeV~\cite{kolbturner}.To this end, 
we calculate the reheating temperature from the total energy density at the time of the last particle's decay:
\be
T_{reh} \backsim \rho_{\textrm{last decay}}^{1/4} \, .
\ee
We find that, in order to get good inflationary parameters, we typically find a low reheating
temperature, e.g., $10^2$-$10^4$~GeV, that resolves the cosmological gravitino problem
while still respecting the nucleosynthesis constraint.

Finally, we consider the prediction of this model for the non-Gaussianity parameter $f_{NL}$.
For this model to be consistent with the Planck results, it needs to predict a value in the range $f_{NL} = 2.7 \pm 5.8$. 
This parameter depends on the sequence of oscillations and decays in the model, and is evaluated at the 
time of the last decay ~\cite{Sueiro:2012fq,Assadullahi:2007uw,Langlois:2011jt}. 
In the model with two inflatons and one curvaton, 
we have seen that the energy density of the heavier inflaton is insignificant from the end of inflation onwards. 
We also note that for such a model to predict a low value of the tensor-to-scalar ratio that is in agreement 
with the Planck constraint, the curvaton must dominate the energy density when it decays. 
Therefore, after the end of inflation, we have the equivalent of a model with one inflaton and one curvaton,
with the curvaton energy density being the dominant component of the energy density of the Universe, i.e., 
$f_\sigma \backsim 1$ at the epoch of last decay. It was shown in \cite{Sueiro:2012fq} that models with one inflaton and 
one curvaton, with both particles contributing to the perturbations, cannot predict a large $f_{NL}$ when the 
curvaton dominates before its decay. In Fig.~2 of \cite{Sueiro:2012fq} we see that, for the range of curvaton expectation 
values that we have in our model and for $f_\sigma \backsim 1$, the non-Gaussianity parameter is small, 
$f_{NL} \in [-1,0]$. So this model with two inflatons and one curvaton is in agreement with the 
constraint imposed by the Planck data.

\section{Three-Sneutrino Inflation}

As already mentioned, part of our motivation for studying extended models of quadratic
inflation was the possibility of rescuing sneutrino inflation. In most supersymmetric
seesaw models for the light neutrino masses, there are three heavy singlet supersymmetric partners of
(right-handed) neutrinos~\footnote{The neutrino oscillation data require only two light neutrinos
to have non-zero masses, which is possible in principle with just two heavy singlet neutrinos in a Type-I seesaw model.}, 
each with a quadratic effective potential. In this Section we 
investigate the possibility of identifying the inflatons and the curvaton of the previous 
Section with these sneutrinos. 
Typical masses of the singlet (right-handed) neutrino
fields are of order $10^8$-$10^{13}$~GeV, without extreme fine-tuning of the Yukawa 
couplings~\footnote{In the Standard Model these range 
from order unity for the top quark down to order $10^{-6}$ for the electron, so a wide range of possible values could be
considered.}. We look here for models that can explain neutrino masses
and mixing angles as well as provide good parameters for Planck observables. 

\begin{figure}[!htbp]
\centering
\begin{tabular}{cc}
\includegraphics[width=80mm, height=80mm]{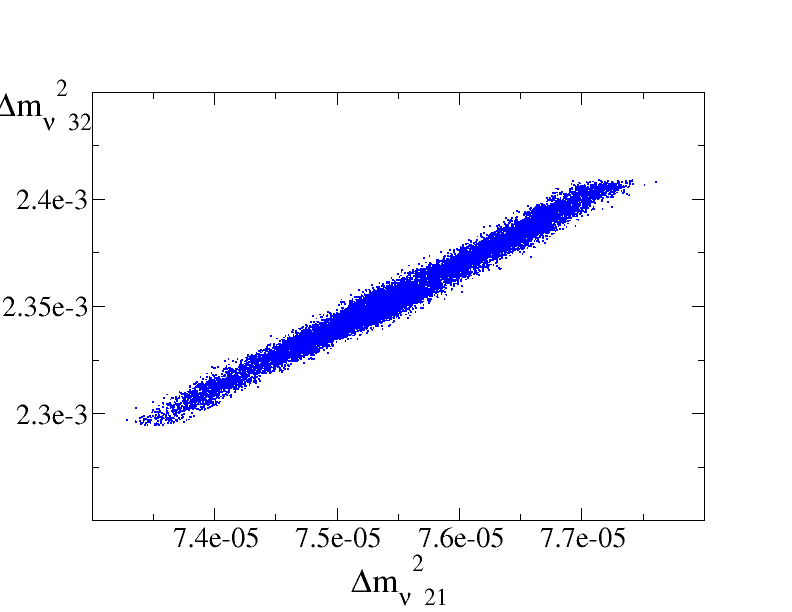}&
\includegraphics[width=80mm, height=80mm]{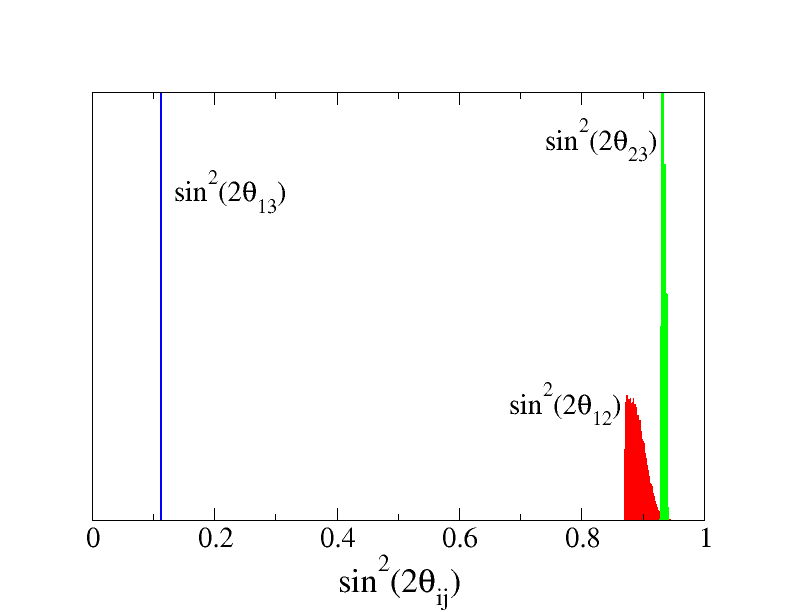}\\
\end{tabular}
\caption{\it In the left panel we show the mass-squared differences for the neutrino model outlined in
the text, and in the right panel we show the three mixing angles. These come from a mass matrix that 
also yields a good fit to Planck data.}
\label{neut}
\end{figure}

We start with the following mass matrix for the singlet (right-handed) (s)neutrinos~\footnote{Supersymmetry
breaking is not important for our analysis.}:
\be
M_{\nu}=\left(
\begin{array}{ccc}
M_\phi & 0 & 0\\
0 & M_\chi & 0\\
0 & 0 & M_\sigma
\end{array}
\right) \, ,
\ee
and a $3\times3$ Yukawa matrix:
\be 
Y_{\nu}=\left(
\begin{array}{ccc}
y_{11}&y_{12}&y_{13}\\
y_{21}&y_{22}&y_{23}\\
y_{31}&y_{32}&y_{33}
\end{array}
\right) \, .
\ee
The decay rate of each heavy particle is then given by ~\cite{Ellis:2004hy}
\be
\Gamma_i=\frac{1}{8\pi}[Y_\nu^\dag Y_\nu]_{ii}M_i \, ,
\ee
where $i=\phi,\chi,\sigma$.

The usual left-handed neutrino mass matrix depends upon the masses and Yukawa coupllings of the heavy
neutrinos, and is given by:
\begin{equation}\label{ssf}
m_{\nu} = v^2 Y^\dag_{\nu}\frac{1}{M_{\nu}} Y_{\nu} \, .
\end{equation}
By diagonalizing this expression we obtain three eigenvalues, which are the three neutrino masses, 
and three eigenvectors, which determine the unitary mixing matrix $U$:
\be
\left(\begin{array}{ccc}
m_{\nu 1} & 0 & 0\\
0 & m_{\nu 2}& 0\\
0 & 0& m_{\nu 3}
\end{array}\right) \; \equiv \; U^\dag m_\nu U \, .
\ee
Here $U$ is the leptonic equivalent of the CKM matrix in the 
quark sector~\cite{nakamura,Rodejohann:2008xp}, which can be written in the following 
parametrization:
\be
U=\left(
\begin{array}{ccc}
c_{12}c_{13} & s_{12}c_{13} & s_{13}\\
-s_{12}c_{23}-c_{12}s_{23}s_{13}& c_{12}c_{23}-s_{12}s_{23}s_{13} & s_{23}c_{13}\\
s_{12}s_{23}-c_{12}c_{23}s_{13}& -c_{12}s_{23}-s_{12}c_{23}s_{13} & c_{23}c_{13}\\
\end{array}\right) \, ,
\ee
where $c_{ij}=\cos(\theta_{ij})$ and $s_{ij}=\sin(\theta_{ij})$.  This matrix contains three neutrino 
mixing angles $\theta_{12}, \theta_{23}, \theta_{13}$. In general, the mixing matrix $U$ can also 
contain CP-violating phases, one detectable in neutrino oscillations and two Majorana phases
that would affect neutrinoless double-$\beta$ decay. For simplicity, here we discard these phases
and assume a real Yukawa matrix for the sneutrinos.

We use the 3 masses $m_{\nu i}$ to fit the two measured light-neutrino mass-squared differences 
$\Delta m^2_{21} = (7.59 \pm 0.20)\times 10^{-5}eV^2$ and $\Delta m^2_{32} = (2.43 \pm 0.13)\times 10^{-3}eV^2$. 
We then choose $U$ to fit the three mixing angles $\sin^2(2\theta_{12})=0.87\pm0.03$, 
$\sin^2(2\theta_{23})>0.92$ and $\sin^2(2\theta_{13})=0.092\pm0.021$, as measured in neutrino oscillation experiments ~\cite{Cheung:2011ph}.

Within this framework, we display one illustrative model with two sneutrino inflatons and one curvaton
that leads to predictions for the tensor-to-scalar ratio and the scalar spectral index that are
consistent with the Planck results. The parameters of this model are outlined in Table~\ref{param},
and the corresponding point in parameter space is shown as a black cross in the right panel of Fig.~\ref{neut}.

\begin{table}
\begin{center}
\begin{tabular}{ | c | c | }
  \hline
Parameter & Value \\                       
\hline
\hline
  $m_\phi$ & 2.6$\times10^{12}$ GeV \\
  $\Gamma_\phi$ & 2.2$\times10^{8}$ GeV \\
$\phi_*$ & 2.2$\times 10^{19}$ GeV \\
\hline
  $m_\chi$ & 2.2$\times 10^{10}$ GeV \\
  $\Gamma_\chi$ & 1.5$\times 10^{4}$ GeV \\
$\chi_*$ & 2.4$\times 10^{19}$ GeV \\
\hline
  $m_\sigma$ & 980 GeV \\
  $\Gamma_\sigma$ & 1.8$\times 10^{-14}$ GeV \\
 $\sigma_*$ & 1.2$\times 10^{17}$ GeV \\
\hline
$T_{reheat}$ & 2.7 $\times 10^3$ GeV \\
$\delta_H$ & 1.8 $\times 10^{-5}$ \\
$n_s$ & 0.9592 \\
$r_T$ & 0.042 \\
\hline
$m_{\nu 1}$ & 4.85$\times 10^{-2}$ eV \\
$m_{\nu 2}$ & 4.92$\times 10^{-2}$ eV \\
$m_{\nu 3}$ & 3.1$\times 10^{-7}$ eV \\
$\theta_{ij}$ & see Fig.~6\\
$\delta m^2_{ij}$ & see Fig.~6 \\
  \hline  
\end{tabular}
\end{center}
\caption{Example of a fit to the Planck and neutrino data. Variations of the parameters around
this particular solution of a few \% are also compatible with the data, and other solutions may exist.
\label{param}}
\end{table}

\section{Conclusions}

In this paper we have looked for the simplest possible fit to the Planck data using only 
fields with quadratic potentials.  We first recalled the well-known result that single-field 
quadratic inflation is under pressure from the Planck data.  We then went on to show that, 
while quadratic inflation with a quadratic curvaton fits the data slightly better, 
it also is disfavoured in its simplest form. This is because the power spectrum is fixed
during the normal single-field inflationary phase, and requiring enough efolds forces 
one to focus on the region where $n_s$ is too high.  We went on to show that with three 
quadratic potentials we can fit both $n_s$ and $r_T$. We also found that such a model provides a minimum in the prediction for the value of $r_T$. Finally, we exhibited a model 
where the three quadratic fields are identified with three sneutrino fields,
two playing the r\^oles of inflatons and one being a curvaton,
displaying one example of a point in parameter space that fits both the neutrino
and cosmological data.

Our results show that it is possible to rescue quadratic inflation, and that this does
not require a very exotic model. Indeed, the three fields required can be identified with
singlet (right-handed) sneutrinos.

\vspace{1cm}
\noindent{ {\bf Acknowledgments} } \\
\noindent
The authors would like to thank David Mulryne for his very useful comments on the first version of this paper.
The work of J.E. was supported in part by
the London Centre for Terauniverse Studies (LCTS), using funding from
the European Research Council 
via the Advanced Investigator Grant 267352.
J.E. and M.F. are grateful for funding from the Science and Technology Facilities Council (STFC).

\end{document}